\documentstyle[aps,preprint,tighten,epsf,floats]{revtex}

\newcommand{\beq}{\begin{equation}}
\newcommand{\eeq}{\end{equation}}
\newcommand{\bqa}{\begin{eqnarray}}
\newcommand{\eqa}{\end{eqnarray}}

\def\square{\vcenter{\vbox{\hrule height.4pt
          \hbox{\vrule width.4pt height8pt
          \kern8pt\vrule width.4pt}\hrule height.4pt}}}

\begin{document}
\draft

\title{Nonuniversal Effects in the Homogeneous Bose Gas}

\author{Eric Braaten$^{a,}$\footnote{braaten@mps.ohio-state.edu}, 
        H.-W. Hammer$^{a,}$\footnote{hammer@mps.ohio-state.edu}, 
        and Shawn Hermans$^{b,}$\footnote{schermans@csbsju.edu} }

\address{$^a$~Department of Physics, The Ohio State University, 
Columbus, OH 43210, USA
~\\$^b$~St. John's University, Collegeville, MN 56321}

\date{December 4, 2000}

\maketitle

\begin{abstract}
Effective field theory predicts that the leading nonuniversal effects
in the homogeneous Bose gas arise from the effective range for 
S-wave scattering and from an effective three-body contact interaction.
We calculate the leading nonuniversal contributions to the energy 
density and condensate fraction and compare the predictions with results 
from diffusion Monte Carlo calculations by Giorgini, Boronat, and Casulleras.
We give a crude determination of the strength of the three-body contact 
interaction for various model potentials.
Accurate determinations could be obtained from diffusion 
Monte Carlo calculations of the energy density with higher statistics.
\end{abstract}

\thispagestyle{empty} 

\newpage

\section{Introduction}

The Bose-Einstein condensation of trapped atoms allows the experimental study
of cold Bose gases with high precision.  It is well-known that the dominant effect
of the interactions between the atoms can be captured by a single constant $a$
called the S-wave scattering length.  This property is often called {\it
universality}.  However, sufficiently accurate measurements will reveal the
sensitivity to aspects of the interatomic interactions other than the
scattering length.  These are called {\it nonuniversal} effects.

Theoretical investigations of the homogeneous Bose gas in the 1950's showed
that its properties could be calculated using a low-density expansion in powers
of $\sqrt {na^3}$, where $n$ is the number density.  For example, the energy
per particle has the expansion
\beq
\label{eden}
{E \over N} = {2 \pi \hbar^2 \over m} a n 
\left \{ 1 + {32 \over 15 \pi} \sqrt {16 \pi n a^3}
+ \left [ {4 \pi - 3 \sqrt 3 \over 6 \pi} \log (na^3) + 
c' \right ] (16 \pi n a^3) + \dots \right \}\,.
\eeq
The leading term is the universal mean-field contribution 
first determined by Bogoliubov \cite{Bog}.  
The $\sqrt {na^3}$ correction was first calculated by Lee, Huang,
and Yang for bosons interacting through a hard-sphere potential
\cite{LeY57}. In the $na^3$ correction, the coefficient of $\log (na^3)$ was 
first calculated by Wu for a hard-sphere potential \cite{Wu59}.
These leading terms in the expansion were shown to be universal 
\cite{univ}, applying to bosons interacting through any short-range 
potential with scattering length $a$.
On the other hand, Hugenholtz and Pines showed that the
constant $c'$ in (\ref{eden}) as well as higher-order 
corrections are not universal \cite{HuP59}.  
They depend on properties of the interactions between the
bosons other than the S-wave scattering length, although the specific
properties were not identified.

Giorgini, Boronat, and Casulleras have studied the properties of the ground
state of homogeneous Bose gases numerically using a diffusion Monte Carlo
method \cite{GBC99}.  
They considered 4 different two-body potentials as models for the
interaction between the bosons,
and calculated the energy per particle and condensate fraction for the
homogeneous gas. 
In the case of the energy per particle, the universal $\sqrt {na^3}$ 
correction in (\ref{eden}) accounts for most of the deviations
from the universal mean-field prediction at the
densities studied, which ranged from $na^3 = 10^{-6}$ to $na^3 = 0.244$.  
However, they also observed small differences between the model potentials, 
i.e. nonuniversal effects.

Effective field theory provides a powerful method for analyzing the
nonuniversal effects in the homogeneous Bose gas.  It identifies precisely
which aspects of the interactions between low-energy bosons  
enter at each order in the expansion in $\sqrt {na^3}$.  
For example, the coefficient $c'$ in the $na^3$
correction in (\ref{eden}) is determined by an
effective three-body contact interaction between low-energy 
bosons \cite{BrN97}.
The $(na^3)^{3/2}$ correction is completely determined by $c'$
and by the effective range for S-wave scattering.

The nonuniversal corrections in the low-density expansion of the Fermi gas
at zero temperature have been studied previously 
\cite{Efimov:1,Efimov:2,fermi,H-F}.
The low-density expansion can be expressed as an expansion in $k_F$,
where $k_F$ is the Fermi wavenumber.
For fermions with a single spin state, the leading correction 
to the energy of an ideal Fermi gas is proportional to $k_F^3  a_p^3$, 
where $a_p$ is the P-wave scattering length \cite{Efimov:1}.
For fermions with $g > 1$ degenerate spin states, the universal corrections 
to the energy can be expanded in powers of $k_F a$,
where $a$ is the S-wave scattering length \cite{Efimov:1}.
The leading nonuniversal corrections are the $k_F^3 a_p^3$ term
and a $k_F^3 a^2 r_s$ term, where $r_s$ is the effective 
range for S-wave scattering.  At order $k_F^4 a^4$,
there is also a nonuniversal correction from the 
effective three-body contact interaction \cite{Efimov:2}.
These corrections were recently discussed within the effective field theory 
framework by Hammer and Furnstahl \cite{H-F}.

In this paper, we use effective field theory to calculate the leading and
next-to-leading nonuniversal corrections to the energy density and condensate
fraction for a homogeneous Bose gas.  
We compare the calculations with the diffusion Monte Carlo 
data from Ref. \cite{GBC99} and attempt to determine
the coefficient of the three-body contact interaction for each of the 
interaction potentials that were considered.

In Section \ref{sec2}, we describe the general philosophy of effective 
theories for dealing with the low-energy behavior of physical systems.  
We use an effective
quantum field theory to identify those parameters that give the leading
nonuniversal contributions to the properties of a homogeneous Bose gas.
The calculations of the leading and next-to-leading nonuniversal contributions 
to the energy density and the condensate fraction are presented in 
Section \ref{sec3}.  
In Section \ref{sec4}, we describe the model interaction potentials used
by Giorgini, Boronat, and Casulleras \cite{GBC99} 
and review their diffusion Monte Carlo results. 
In Section \ref{sec5}, we analyze the diffusion Monte Carlo results
of Ref. \cite{GBC99} and give a crude determination of the  
three-body contact parameter for each of the potentials that was considered.
A summary and our conclusions are presented in Section \ref{sec6}.

\section{Effective Field Theory}
\label{sec2}

Effective theory is a general approach to understanding the low energy behavior
of a physical system that has deep roots in several branches of physics.  Some
of these roots are described in Ken Wilson's Nobel lecture on the
Renormalization Group \cite{Wil83}.  In elementary particle physics, the roots
have two main branches. One branch evolved from efforts
to make intuitive sense of
renormalization in quantum electrodynamics (QED) \cite{Lep89}.  
In perturbative calculations in QED, intermediate steps are
plagued by \lq\lq ultraviolet divergences" that indicate strong sensitivity to
physics at extremely short distances.  The renormalization procedure 
eliminates the ultraviolet divergences and allows
extremely accurate predictions of the properties of electrons and their
bound states in terms of two fundamental parameters:  the electron mass and the
fine structure constant.  The other main branch evolved from 
efforts to understand
the low-energy behavior of pions and nucleons.  The fundamental description of
these strongly interacting particles is provided by quantum chromodynamics
(QCD).  However, an effective theory can be used to describe the low-energy
behavior of these particles accurately without using any
information about QCD other than its global symmetries and measurements of a
few low-energy observables \cite{Wei79}.  
Starting from these two main roots, effective field theory has developed into a
universal language for modern elementary particle physics \cite{eftrefs}.
It is used to develop low-energy approximations to the Standard
Model of elementary particle physics and also to quantify corrections to
the Standard Model arising from a more 
fundamental theory, such as a unified field theory or string theory.

Most of the applications of effective theories to date have been carried out
within the context of quantum field theory.  However, the principles of
effective theory apply equally well to the simple quantum mechanics
problem of two particles interacting through a central potential $U(r)$ 
\cite{Lep97}. Suppose we are interested only in the low-energy
observables of the system, where ``low-energy'' refers to energy 
$E$ close to the threshold $E=0$.  
These observables include bound-state energy levels close
to threshold and the low-energy scattering amplitudes.  Suppose also that we
know the potential $U(r)$ accurately at long distances $r > R$, but that
its behavior at short distances $r < R$ is not known accurately.  
(If it is a
short-range potential with range smaller than $R$, then $U(r) = 0$  
for $r >R$.  If the particles are real atoms interacting at long distances 
through a van der Waals potential, then $U(r) = -C_6 / r^6$ for $r > R$.)  
Given sufficiently precise information about some low energy observables, 
effective theories allow
other low-energy observables to be calculated with arbitrarily high
accuracy without having any information about the short-distance potential.

The method is very simple.  Simply replace $U(r)$ by an effective
potential $U_{\rm eff}(r)$ that is identical for $r > R$ and whose form 
for $r <R$
involves an adjustable parameter $c_1$.  For $r <R$, the effective potential
need not bear any resemblance to the original potential $U(r)$ as long 
as it has an adjustable parameter. A sufficient condition on the short-distance
potential is that by varying $c_1$, one should be able to change the number 
of bound states by $\pm 1$. This guarantees that the scattering length $a$
can be tuned to any value from $-\infty$ to $+\infty$. A simple example
is a constant potential with adjustable depth $V_1$.
Tune the value of this parameter $c_1$ so that
the scattering amplitude at threshold is reproduced exactly.  
Then the Schr{\"o}dinger equation with $U_{\rm eff} (r)$ 
will reproduce all other low-energy observables with errors that scale
linearly with $E$.  To achieve higher accuracy, 
use an effective potential $U_{\rm eff}(r)$ with two independent 
adjustable parameters $c_1$ and $c_2$.
A simple example is a 2-step potential with adjustable depths $V_1$ and
$V_2$.
The equality of $U_{\rm eff}(r)$ and $U(r)$ for $r>R$
guarantees that the difference between their scattering amplitudes 
is an analytic function of the wavevector ${\bf k}$. 
If we tune $c_1$ and $c_2$ so that the S-wave scattering amplitudes 
agree to second order in $k=|\bf k|$, then $U_{\rm eff}(r)$ will reproduce
all low-energy S-wave observables with errors that scale like $E^2$.  
By tuning a third parameter $c_3$, we can guarantee that P-wave
observables also have errors that scale like $E^2$. 
If we add more adjustable parameters and tune
the scattering amplitude to higher order in $E$, the errors will scale as
higher powers of $E$.  Low-energy observables at
energies within the appropriate radius of convergence can therefore be 
calculated to any specified accuracy by adjusting 
a finite number of parameters in the effective potential.

The problem of a Bose gas at zero temperature is an ideal application of
effective field theory.  Suppose we have a gas with number density $n$
consisting of identical bosons of mass $m$ interacting through a short-range
two-body potential $U({\bf r})$.  (For convenience, we will refer to these
particles as ``atoms.")  Using the method of second quantization, this system
can be described by a quantum field theory.  The quantum field operator 
$\psi({\bf r}, t)$ annihilates an atom at the point $\bf r$, 
and $\psi^\dagger \psi$ is the number density operator. 
The dynamics of this quantum field theory can be
summarized by a nonlocal Lagrangian density:
\bqa
{\cal L} & = & {1 \over 2} i \hbar 
\left[ \psi^* {\dot \psi} -  {\dot \psi}^* \psi \right]
- {\hbar^2 \over 2m} \nabla \psi^* \cdot \nabla \psi + \mu \psi^* \psi
- {1 \over 2} \int d^3 r' \psi^* \psi({\bf r}) 
	U({\bf r} - {\bf r'}) \psi^* \psi({\bf r'}) \,.
\label{Lnonloc}
\eqa
The parameter $\mu$ is the chemical potential. 
If we wish to describe a system consisting of a small number of atoms, 
such as a two-body or three-body system, then the $\mu \psi^* \psi$ 
term can be ignored.  
To describe a homogeneous gas of number density $n$, 
$\mu$ must be tuned so that the expectation value of $\psi^\dagger \psi$ 
in the ground state of the quantum field theory is equal to $n$.

Now suppose we are interested only in low-energy observables, such as the
properties of the ground state.  The principles of effective theory tell us
that the short-range potential $U({\bf r})$ can be replaced by any other set of
effective short-range interactions, so long as they have adjustable parameters
that can be tuned to reproduce a set of low-energy measurements.  An example of
such a set of low-energy measurements is the coefficients of the expansions in
$\sqrt {na^3}$ for the energy density and other properties of the many-body
system.  A simpler example is the coefficients in the low-momentum expansions 
for the scattering amplitudes of atoms in the two-body sector, 
three-body sector, etc..

If we wish to carry out analytic calculations, one particularly convenient
choice for the effective short-range interactions is to take them all to be
purely local.  The dynamics of the resulting quantum field theory is then
described by a local Lagrangian density:
\bqa
\label{Lloc}
{\cal L}_{\rm eff} & = & {1 \over 2} i \hbar 
\left[ \psi^* {\dot \psi} -  {\dot \psi}^* \psi \right]
- {\hbar^2 \over 2m} \nabla \psi^* \cdot \nabla \psi + \mu \psi^* \psi \\
&& -\frac{A_0}{4} ( \psi^* \psi)^2 - \frac{A_2}{4} \nabla(\psi^* \psi)
\cdot  \nabla\left(\psi^* \psi\right)- \frac{B_0}{36} (\psi^* \psi)^3 
+ \dots \nonumber
\eqa
The terms in the Lagrangian are constrained by the symmetries of the original
Lagrangian:  Galilean invariance, parity, and time reversal.  
We have included explicitly those terms that contribute 
the leading nonuniversal corrections to the properties of the homogeneous gas.
All other terms are either higher order in $\psi$ or in $\nabla$.  
The $[\nabla(\psi^* \psi)]^2$ term is the only possible term 
that is fourth order in $\psi$ and second order in $\nabla$.
Up to total divergences, the only other independent term is
$(\psi^* \nabla \psi - \psi \nabla \psi^*)^2$, 
but it is forbidden by Galilean invariance. Note that while the original
Lagrangian (\ref{Lnonloc}) included only terms of 2nd and 4th order in $\psi$, 
the effective Lagrangian  (\ref{Lnonloc}) 
includes terms of 6th order and higher.  The 6th order terms are
necessary, because a local Lagrangian without such terms could not reproduce
the low-energy behavior of scattering  amplitudes in the three-body sector.  
It would also be unable to reproduce the $n a^3$ 
correction term in the low-density
expansion (\ref{eden}) for the energy density of the homogeneous gas.

The coefficients $A_0$ and $A_2$ in (\ref{Lloc}) can be determined 
by demanding that the effective theory reproduces the amplitude for 
low-energy atom-atom scattering up to errors that scale like the
square of the energy.
The T-matrix element at tree-level for two atoms with 
wavenumbers ${\bf k_1}, {\bf k_2}$ to scatter into atoms with 
wavenumbers ${\bf k_1'}, {\bf k_2'}$
can be deduced from the interaction terms in (\ref{Lloc}):
\bqa
\label{Ttree}
\left( {\cal T}_2 \right)_{\rm tree} &=& - A_0 
+ {1 \over 2} A_2 
\left[ ({\bf k_1}-{\bf k_1'}) \cdot ({\bf k_2}-{\bf k_2'}) + 
	({\bf k_1}-{\bf k_2'}) \cdot ({\bf k_2}-{\bf k_1'}) \right]  \,.
\eqa
Going to the center of momentum frame with 
${\bf k_1} = - {\bf k_2} = {\bf k}$ and
${\bf k_1'} = - {\bf k_2'} = {\bf k'}$ 
and using the energy conservation condition $k^2 = {k'}^2$, 
this reduces to 
\bqa
\left( {\cal T}_2 \right)_{\rm tree} &=& - A_0 - 2 A_2 k^2  \,.
\eqa
The corrections from loop diagrams depend on the regularization and 
renormalization prescriptions, but they form a geometric series 
that can be summed analytically.
If we impose an ultraviolet cutoff $\Lambda$ on the wavenumber,
the exact T-matrix element 
can be determined analytically by solving a simple integral equation and
has the form
\bqa
{\cal T}_2 &=& -
\left[  {1 \over  {\cal A}_0 + 2 {\cal A}_2 k^2} + 
{m \over 4 \pi^2 \hbar^2}\left( \Lambda+
	i {\pi \over 2} k\right) \right]^{-1}  \,,
\label{T2-cutoff}
\eqa
where ${\cal A}_0$ and ${\cal A}_2$ are complicated functions of 
$A_0$, $A_2$, and $\Lambda$.
The absence of any angular dependence indicates that there is
S-wave scattering  only.
We get a much simpler expression if we use dimensional regularization,
which automatically subtracts ultraviolet divergences that grow like a 
power of $\Lambda$. In this case the exact T-matrix element is
\bqa
{\cal T}_2 &=& -
\left[  {1 \over  A^{({\rm dr})}_0 + 2 A^{({\rm dr})}_2 k^2} 
	+ i {m k \over 8 \pi \hbar^2} \right]^{-1}  \,.
\label{T2-dimreg}
\eqa
The expressions (\ref{T2-cutoff}) and (\ref{T2-dimreg}) for ${\cal T}_2$
do not have precisely the same functional dependence on $k$.
However, if we make the identifications
$1/A^{({\rm dr})}_0=1/{\cal A}_0+m\Lambda/(4\pi\hbar^2)$ 
and $A^{({\rm dr})}_2/(A^{({\rm dr})}_0)^2={\cal A}_2/({\cal A}_0)^2$,
the expansions of (\ref{T2-cutoff}) and (\ref{T2-dimreg})
for small $k$ are identical through order $k^3$. The difference scales like
$E^2$, where $E=\hbar^2 k^2/m$ is the total energy. From the effective
theory point of view, including the $A_2$ term in the effective Lagrangian
(\ref{Lloc}) allows low-energy observables in the two-body sector to
be reproduced up to errors that scale like $E^2$. 
Different regularizations require different values of $A_0$ and $A_2$,
but they give the same low-energy observables up to errors that
scale like $E^2$.

The amplitude for atom-atom scattering in the original theory described
by the Lagrangian (\ref{Lnonloc}) can be 
determined by solving the Schr{\"o}dinger 
equation with the appropriate interatomic 
potential $U({\bf r}-{\bf r'})$.  
The contribution from S-wave scattering can be 
expressed in a manifestly unitary form in terms of the S-wave
phase shift $\delta_0(k)$:
\bqa
{\cal T}_2 &=& 
{8 \pi \hbar^2 \over m k} e^{i \delta_0(k)} \sin[\delta_0(k)] \,.
\label{T2-ps}
\eqa
The scattering length $a$ and the effective range $r_s$ for S-wave scattering 
are defined by the low-momentum expansion of the phase shift.
Because of the identity $e^{i \delta} \sin \delta = 1/(\cot \delta - i)$,
the low-momentum expansion is conveniently written in the form
\bqa
\label{ERexp}
k \cot[\delta_0(k)] &=& 
- {1 \over a} + {1 \over 2} r_s k^2 + \ldots \,, 
\eqa
which defines the effective range $r_s$.
After inserting this into (\ref{T2-ps}), 
the T-matrix element can be expanded in powers of $k$.
The expansions of (\ref{T2-dimreg}) and (\ref{T2-ps}) match through 
third order in $k$ if the coefficients  
$A_0^{({\rm dr})}$ and $A^{({\rm dr})}_2$ have the values
\bqa
\label{A0match}
A^{({\rm dr})}_0 &=& \frac{8\pi\hbar^2 a}{m}\,,
\\
\label{A2match}
A^{({\rm dr})}_2 &=& \frac{8\pi\hbar^2 a}{m} \times \frac{ar_s}{4} \,.
\eqa
These are the coefficients if we use dimensional regularization 
to remove ultraviolet divergences.
If we used a wavenumber cutoff as in (\ref{T2-cutoff}),
the coefficients $A_0$ and $A_2$ in the Lagrangian would depend on 
$\Lambda$ in a very complicated way.
However, their values are uniquely determined by the dimensional
regularization coefficients (\ref{A0match}) and (\ref{A2match}).

The $(\psi^* \psi)^3$ term in the Lagrangian (\ref{Lloc})
is necessary in order for the effective theory to reproduce 
the amplitude for the low-energy scattering of three atoms with errors
that scale linearly in the energy $E$. At tree level, this term
gives a momentum independent contribution $-B_0$ to the T-matrix element 
${\cal T}_3$ for $3\to 3$ scattering. 
One can see the necessity for such a term by considering the 
perturbative expansion for ${\cal T}_3$ in powers of  $A_0$,
or equivalently $a$.  At 4th order in $a$, there are 2-loop 
Feynman diagrams that depend logarithmically on the
ultraviolet cutoff $\Lambda$ for the wavenumber of atoms in 
intermediate states \cite{BrN97}. They give an additional 
momentum-independent contribution to ${\cal T}_3$ that is universal.
The T-matrix element therefore includes the terms
\beq
{\cal T}_3 = -B_0+
{3(4 \pi - 3 \sqrt 3) \over 32 \pi^3} {m^3 A_0^4 \over \hbar^6} \log 
{\Lambda \over \kappa} + \ldots \,,
\label{T-3}
\eeq
where $\kappa$ is a wavenumber set by the energies of the scattering 
particles. In the original theory, the ultraviolet cutoff is provided by the
interaction potential $U({\bf r})$. The ultraviolet divergence
in (\ref{T-3}) therefore indicates that there is a momentum-independent
contribution to $3\to 3$ scattering that depends on the 
interaction potential.
The T-matrix element (\ref{T-3}) can be made independent of the cutoff 
$\Lambda$ if the coefficient $B_0$ depends on $\Lambda$
as specified by the following renormalization group equation:
\beq
\label{rg1}
\Lambda {d \ \over d \Lambda} B_0 
= {3 (4 \pi - 3 \sqrt{3}) \over 32 \pi^3}  {m^3 A_0^4 \over \hbar^6}\,.
\eeq
The parameter $B_0$ could in principle be determined 
from the numerical solution to the three-body problem for the scattering
of three atoms interacting pairwise through the potential $U({\bf r})$.
However, it can equally well be determined from calculations of the 
ground state energy for a large number of atoms interacting 
through the potential $U({\bf r})$.  
This is the strategy we will follow in this paper. 

Using dimensional analysis, the coefficient of the three-body contact 
interaction term in (\ref{Lloc}) can be written as
\beq
\label{B0match}
B_0=\frac{8\pi\hbar^2 a^4}{m} \times (144 \pi c) \,,
\eeq
with $c$ a dimensionless constant. The factor of $144\pi$ has been chosen 
for later convenience. We will refer to $c$ as the three-body contact
parameter.

\section{Nonuniversal corrections}
\label{sec3}

In this section, we calculate the leading and next-to-leading 
nonuniversal corrections to the energy density of a homogeneous Bose gas. 
We also calculate the leading nonuniversal corrections to the 
condensate fraction.

\subsection{Mean-field corrections to the energy density}

The leading nonuniversal correction to the energy density ${\cal E}$ can be
obtained by applying the mean-field approximation to the Lagrangian ${\cal
L}_{\rm eff}$ in (\ref{Lloc}), including the $(\psi^* \psi)^3$ term.  In the
mean-field approximation, $\psi$ is replaced by its ground-state expectation
value $v$, which can be taken real-valued. 
The condensate number density is $n_0 = v^2$.
The free energy density ${\cal F}$ in the mean-field approximation 
is simply $- {\cal L}_{\rm eff}$, with $\psi$ replaced by $v$:
\beq
\label{Fmf}
{\cal F}_{\rm mf} = - \mu v^2 + {1 \over 4} A_0 v^4 + {1 \over 36} B_0 v^6 \,.
\eeq
Setting $v^2 = n_0$ in the energy density ${\cal E} = {\cal F}+\mu n$,
it becomes
\beq
\label{Emf1}
{\cal E}_{\rm mf} = {1 \over 4} A_0 n_0^2 + {1 \over 36} B_0 n_0^3 
	+ \mu \, (n-n_0)\,.
\eeq
The chemical potential is obtained from the condition 
$\partial {\cal F}/\partial v  = 0$, which guarantees that $v$ minimizes 
the free energy:
\beq
\label{mumin}
\mu_{\rm mf} = {1 \over 2} A_0 n_0 + {1 \over 12} B_0 n_0^2\,.
\eeq
Substituting the expressions (\ref{A0match})  and (\ref{B0match})
for the coefficients into (\ref{Emf1}), we have
\beq
\label{E-mf}
{\cal E}_{\rm mf} = {2 \pi \hbar^2 \over m} a n_0^2 
\left[ 1 + c (16 \pi n_0 a^3) \right]  + \mu \, (n-n_0) \,.
\eeq
In the mean-field approximation, 
the number density $n = \psi^* \psi$ reduces to $n = v^2$.
Setting $n = n_0$ in (\ref{E-mf}),  
the first term gives the familiar universal mean-field energy density:
\bqa
{\cal E}_{\rm umf} & = & {2 \pi \hbar^2 \over m} a n^2 \,.
\label{E-umf}
\eqa
The constant $c$  in (\ref{E-mf}) corresponds to a nonuniversal 
contribution to the constant $c'$ 
under the logarithm in  the coefficient of $na^3$ in (\ref{eden}). 

The universal $\log (na^3)$ term in the coefficient of
$na^3$ in (\ref{eden}) can be obtained through complicated two-loop 
calculations in many-body quantum field theory.  
However, as pointed out by Braaten and Nieto
\cite{BrN97}, it follows from the term in (\ref{E-mf}) proportional
to $c$ by a simple renormalization group argument.  
If $3\to 3$ scattering amplitudes are to be independent of the ultraviolet
cutoff $\Lambda$ on the wavenumber of atoms in the intermediate states,
the coefficient $c$ defined by (\ref{B0match}) must depend logarithmically on 
the ultraviolet cutoff $\Lambda$ in the way prescribed by the renormalization
group equation (\ref{rg1}).  
The mean-field contribution (\ref{Emf1}) to the energy density depends on the
ultraviolet cutoff through the coefficient $c(\Lambda)$.  The total energy
density after renormalization must be independent of $\Lambda$.  Therefore,
there must be loop diagrams in many-body perturbation theory that have
logarithmic ultraviolet divergences  that precisely cancel the
$\Lambda$-dependence of $c(\Lambda)$.  
The only momentum scale in these diagrams is
the inverse of the coherence length $\xi = (16 \pi n_0a)^{-1/2}$,
which appears in the quasiparticle dispersion relation. 
Thus the logarithmic divergences are of the form $\log (\Lambda \xi)$.  
The energy density can be independent of $\Lambda$ 
only if the explicit logarithmic
divergences and the coefficient $c(\Lambda)$ appear in the combination
\bqa
&&
c (\Lambda) - {4 \pi - 3 \sqrt{3} \over 3 \pi} 
	\log(\Lambda/\sqrt{16\pi n_0a})
\nonumber 
\\
&&
\;=\; c (\Lambda) - {4 \pi - 3 \sqrt{3} \over 3 \pi} \log(\Lambda a)
+ {4 \pi - 3 \sqrt{3} \over 6 \pi} \log(16\pi n_0a^3) \,.
\label{clog-comb}
\eqa
The three-body contact parameter $c$ need not be defined by a wavenumber cutoff
$\Lambda$.  For example, one can define a $\Lambda$-independent parameter 
$c$ by absorbing the logarithm of $\Lambda a$ in (\ref{clog-comb}) into
$c(\Lambda)$. Alternatively, one can use dimensional regularization and minimal
subtraction to define a parameter $c(\kappa)$ that depends on an adjustable
renormalization scale $\kappa$. One could even define $c$ by making a 
specific choice for the universal constant under the logarithm 
in the low-density expansion for the energy per particle in (\ref{eden}).
We will adopt such a definition later in this paper. 
Regardless of the definition, any dependence of a many-body
observable on $c$ must be accompanied by a logarithm of $n_0a^3$
in the combination
\beq
c + {4 \pi - 3 \sqrt{3} \over 6 \pi} \log(16\pi n_0a^3) \,.
\label{c-comb}
\eeq
Replacing $c$ in (\ref{E-mf}) by this combination
and setting $n_0 = n$, we reproduce the universal logarithmic 
$na^3$ correction in (\ref{eden}).

\subsection{ Semiclassical corrections to the energy density}

The next-to-leading order nonuniversal corrections to the energy density can be
obtained by computing the semiclassical (one-loop) correction to ${\cal E}$,
including the effects of the effective range term and the three-body contact
interaction in ${\cal L}_{\rm eff}$.  The semiclassical correction before
renormalization is simply the sum of the zero-point energies of the
quasiparticles:
\beq
\label{Esc}
{\cal E}_{\rm sc} = \int {d^3 k \over (2 \pi)^3} {1 \over 2}  \epsilon (k) \,,
\eeq
where $\epsilon (k)$ is the quasiparticle dispersion relation.  
This expression is ultraviolet divergent, 
and renormalization is necessary to obtain a finite result.
If we insert the Bogoliubov dispersion relation 
\beq
\epsilon_{\rm Bog} (k) = {\hbar^2 \over 2m} k \sqrt {k^2 + 16 \pi a n_0 } \,,
\label{disp-Bog}
\eeq
renormalization can be accomplished by subtracting powers of $k$ from the
integrand to make it convergent:
\beq
{\cal E}_{\rm LHY} = {\hbar^2 \over 8 \pi^2 m} 
\int^\infty_0 dk \, k^3 \left [ \sqrt
{k^2 + 16 \pi n_0a} - k - 8 \pi n_0a {1 \over k} 
	+ 32 \pi^2 n_0^2 a^2 {1 \over k^3}
\right] \,.
\label{Elhy-1}
\eeq
The three subtractions can be identified with renormalizations of the 
vacuum energy, the chemical potential, and the scattering length, 
respectively.  If we imposed a cutoff $\Lambda$ on the wavenumber,
the three subtractions would be proportional to $\Lambda^5$, $\Lambda^3$,
and $\Lambda$, respectively.  The result of evaluating the convergent 
integral in (\ref{Elhy-1}) is
\beq
{\cal E}_{\rm LHY} = {\hbar^2 \over 60 \pi^2 m} (16 \pi n_0 a)^{5/2} \,.
\label{Elhy-3}
\eeq
Setting $n_0=n$, this reproduces the $\sqrt {na^3}$ correction 
in (\ref{eden}) first calculated by Lee, Huang, and Yang \cite{LeY57}.

The result (\ref{Elhy-3}) can also be obtained by using dimensional 
regularization. 
The integral is generalized from 3 dimensions to $3-2\epsilon$ dimensions,
evaluated for a value of $\epsilon$ for which it converges, and
then analytically continued to $\epsilon = 0$:
\beq
{\cal E}_{\rm LHY} = {\hbar^2 \over 8 \pi^2 m} 
\int^\infty_0 dk \, k^{3-2 \epsilon} \sqrt {k^2 + 16 \pi a n_0} 
\bigg|_{\epsilon \to 0} \,.
\label{Elhy-2}
\eeq
To be more specific, the integral is separated into two regions:
$k<k^*$ and $k>k^*$, where $k^*$ is an arbitrary wavenumber.
The contribution from $k<k^*$ is evaluated for complex values of 
$\epsilon$ satisfying ${\rm Re} \, \epsilon < 2$.
The contribution from $k>k^*$ is evaluated for complex values of 
$\epsilon$ satisfying ${\rm Re} \, \epsilon > {5\over2}$.
The sum of the two contributions is 
\beq
{\cal E}_{\rm LHY} = {\hbar^2 \over 8 \pi^2 m} 
{\Gamma(2-\epsilon) \Gamma(-{5\over2}+\epsilon) \over 2 \Gamma(-{1\over2})}
(16 \pi a n_0)^{{5\over2}-\epsilon} \,.
\label{Elhy-4}
\eeq
Setting $\epsilon = 0$, we recover (\ref{Elhy-3}).

The effective range and three-body contact terms in (\ref{Lloc}) change the
quasiparticle dispersion relation $\epsilon (k)$.  
To determine $\epsilon (k)$, we substitute 
$\psi({\bf r},t) = v + \phi({\bf r},t)$ into (\ref{Lloc}).
The quasiparticle Lagrangian ${\cal L}_2$ consists of the terms 
that are second order in $\phi$. 
Using the mean-field expression (\ref{mumin}) 
for the chemical potential and setting $v^2 =n_0$, 
the quasiparticle Lagrangian reduces to
\bqa
{\cal L}_2 & = & 
{1 \over 2} i \hbar 
\left[ \phi^* {\dot {\phi}} 
	-  {\dot {\phi}}{}^* \phi \right]
- {\hbar^2 \over 2m} 
\left[ \nabla \phi^* \cdot \nabla \phi 
	+ 4 \pi n_0 a^2 r_s \nabla ({\rm Re} \phi) 
		\cdot \nabla ({\rm Re} \phi) \right]
\nonumber 
\\
& & 
- {\hbar^2 \over 2m}(16 \pi n_0 a) [1 + 3 c(16 \pi n_0 a^3)] 
	\left( {\rm Re} \phi \right)^2
\,.
\eqa
The quasiparticle dispersion relation $\epsilon (k)$ can be determined by
Fourier transforming the fields ${\rm Re} \phi$ 
and ${\rm Im} \phi$ and finding the eigenvalues
of the resulting $2 \times 2$ matrix.  The result is 
\beq
\epsilon (k) = {\hbar^2 \over 2m} k \sqrt { (1 + 4 \pi n_0 a^2 r_s) k^2 + 
16 \pi a n_0 [1 + 3 c (16 \pi n_0 a^3)]}\,.
\label{disp-nonu}
\eeq
Setting $r_s = c = 0$, we recover the Bogoliubov 
dispersion relation in (\ref{disp-Bog}).

Since the dispersion relation (\ref{disp-nonu}) has the same functional form 
as the Bogoliubov dispersion relation (\ref{disp-Bog}), 
we can compute the semiclassical energy simply by
rescaling the integrand and the integration variable $k$
in the dimensionally regularized expression (\ref{Elhy-2}):
\beq
{\cal E}_{\rm sc} = {\hbar^2 \over 60 \pi^2 m} (1 + 4 \pi n_0 a^2 r_s)^{-2} 
\left\{ 16 \pi n_0 a [ 1 + 3 c (16 \pi n_0 a^3)] \right\}^{5/2} \,.
\eeq
Expanding to first order in $r_s$ and $c$, this reduces to
\beq
{\cal E}_{\rm sc} = {2 \pi \hbar^2 \over m} a n_0^2 
\left\{ {32 \over 15 \pi} \sqrt {16 \pi n_0 a^3} 
+  \left[  {16 \over \pi}c - {16 \over 15 \pi} {r_s \over a} \right] 
	(16 \pi n_0 a^3)^{3/2} \right \}\,.
\label{E-SC}
\eeq

\subsection{Semiclassical corrections to the number density}

The total number density is $n = \langle \psi^\dagger \psi \rangle$,
while the number density of atoms in the condensate is 
$n_0 = | \langle \psi \rangle |^2$.
In the mean-field approximation, the operator $\psi$ 
is replaced by its ground-state expectation value $v$,
in which case $n=n_0$.
In the semiclassical approximation, we have
\beq
n_{\rm sc} \;=\;
n_0 + {1 \over 4}\int\frac{d^3 k}{(2\pi)^3}
\left[ {\hbar^2 k^2/(2m) \over \epsilon(k)}
	+{\epsilon(k) \over \hbar^2 k^2/(2m) } \right]\,.
\label{n-sc:1}
\eeq
Inserting the Bogoliubov dispersion relation (\ref{disp-Bog})
and using dimensional regularization to remove the 
power ultraviolet divergences, this can be written
\beq
n_{\rm Bog} \;=\;
n_0 + {1 \over 8 \pi^2} \int_0^\infty dk k^{2-2\epsilon} 
\left[ {k \over \sqrt{k^2 + 16 \pi n_0 a}}
	+ {\sqrt{k^2 + 16 \pi n_0 a} \over k} \right]
		\Bigg|_{\epsilon \to 0}\,.
\label{n-Bog:1}
\eeq
Evaluating the integrals and analytically continuing 
to $\epsilon = 0$, we obtain the classic result of 
Bogoliubov \cite{Bog}:
\beq
n_{\rm Bog} \;=\;
n_0 + {1 \over 24 \pi^2} ( 16 \pi n_0 a)^{3/2} \,.
\label{n-Bog:2}
\eeq

In order to compute the complete semiclassical correction to the 
condensate density, including the effects of the effective range 
and the three-body contact coefficient, we must insert the 
quasiparticle dispersion relation (\ref{disp-nonu}) into 
the integral (\ref{n-sc:1}).  Since (\ref{disp-nonu}) 
has the same functional form as the Bogoliubov dispersion relation 
(\ref{disp-Bog}), we can evaluate the integral simply by
rescaling the integrand and the integration variable $k$
in the expression (\ref{n-Bog:1}):
\begin{eqnarray}
\label{n-sc:2}
n_{\rm sc} 
&=& n_0 + {1 \over 24 \pi^2} 
\left[ 2 (1 + 4 \pi n_0a^2 r_s)^{-2} - (1 + 4 \pi n_0a^2 r_s)^{-1} \right] 
\\ & & \hphantom{n_0 + {1 \over 24 \pi^2}}\times
\left\{ 16 \pi n_0 a [ 1 + 3 c (16 \pi n_0 a^3)] \right\}^{3/2} \;.
\nonumber
\end{eqnarray}
Expanding to first order in $r_s$ and $c$, this reduces to
\beq
\label{n-sc:3}
n_{\rm sc} \;=\; n_0 
\left( 1 + {2 \over 3 \pi} \sqrt{16 \pi n_0 a^3}
+ \left[{3 \over \pi}  c - {1 \over 2 \pi} {r_s \over a} \right]
	(16 \pi  n_0a^3)^{3/2} \right) \,.
\eeq

\subsection{Low-density expansion for the condensate fraction}

The semiclassical expression (\ref{n-sc:3}) for the number density
includes both the leading universal corrections and the leading 
nonuniversal corrections to the mean-field result $n = n_0$.
The coefficient $c$ in the $(n_0a^3)^{3/2}$ correction must be 
accompanied by a universal logarithmic correction in the combination
(\ref{c-comb}). There are universal corrections proportional to
$(n_0a^3)$ and $(n_0a^3)^{3/2}$ whose coefficients are as yet unknown.
Including all corrections through order $(n_0a^3)^{3/2}$,
the expression for the number density is
\begin{eqnarray}
n &=& n_0 
\left\{ 1 + {2 \over 3 \pi} \sqrt{16 \pi n_0 a^3}
	+ d (16 \pi n_0 a^3)
\right.
\nonumber
\\ 
&& \hspace{1cm} \left.		
	+ \left( {3 \over \pi} \left[  c_E
		+ {4 \pi - 3 \sqrt 3 \over 6 \pi} 
		\log (16 \pi n_0 a^3)\right]
	- {1 \over 2 \pi} {r_s \over a} 
	+ e \right)
		(16 \pi  n_0a^3)^{3/2} 
\right\}\,.
\label{n-gen}
\end{eqnarray}
The parameters $c_E$ and $r_s$ depend on the interaction potential,
while $d$ and $e$ are universal coefficients.
The subscript on $c_E$ is a reminder that we will define 
this parameter by the form of the low-density expansion for the energy
of the homogeneous gas. 

The low-density expansion for the condensate fraction $n_0/n$
can be obtained by inverting (\ref{n-gen}) to obtain $n_0$ as
a function of $n$.  Including all terms through order $(na^3)^{3/2}$,
the expression for the condensate fraction is
\bqa
{n_0 \over n} &=& 
1 - {2 \over 3 \pi} \sqrt{16 \pi n a^3}
	- d' (16 \pi n a^3)
\nonumber
\\ 
&& \hspace{1cm} 
	- \left( {3 \over \pi} \left[  c_E
		+ {4 \pi - 3 \sqrt 3 \over 6 \pi} 
		\log (16 \pi n a^3)\right]
	- {1 \over 2 \pi} {r_s \over a} 
	+ e' \right) 
	(16 \pi  na^3)^{3/2}\,,
\label{n0-gen}
\eqa
where $d' = d - {2\over3\pi^2}$ and 
$e' = e - {7\over3\pi}d + {7\over9\pi^3}$.

\subsection{Low-density expansion for the energy density}

The mean-field contribution to the energy density in (\ref{E-mf}) 
includes the leading nonuniversal correction 
to the universal mean-field energy density.
The semiclassical contribution in (\ref{E-SC})
includes the leading universal correction
and the next-to-leading nonuniversal corrections. 
The coefficient $c$ in the nonuniversal correction must be 
accompanied by a universal logarithmic correction in the combination
(\ref{c-comb}). There is a universal correction proportional 
to $n_0 a^3$ that was calculated by Braaten and Nieto \cite{BrN99}.
There is also a universal correction proportional 
to  $(n_0 a^3)^{3/2}$ that has not been calculated.  
Including all corrections through order $(n_0a^3)^{3/2}$,
the expression for the energy density is
\bqa
{\cal E} &=&
{2 \pi \hbar^2 \over m} a n_0^2  
\left\{ 1 + {32 \over 15 \pi} \sqrt {16 \pi n_0 a^3}
	+ \left [ c_E + {4 \pi - 3 \sqrt 3 \over 6 \pi} 
			\log (16 \pi n_0 a^3) 
	+ {4 \over 9\pi^2} \right ] (16 \pi n_0 a^3) 
\right.
\nonumber
\\
&&  \hphantom{{2 \pi \hbar^2 \over m} a n_0^2 \{ } \left.
+ \left( {16 \over \pi} \left[ c_E 
	+ {4 \pi - 3 \sqrt 3 \over 6 \pi} 
		\log (16 \pi n_0 a^3) \right]
	- {16 \over 15\pi} {r_s \over a} 
	+ b \right) 
	(16 \pi n_0 a^3)^{3/2} \right\}  
\nonumber
\\
&&  + \mu \, (n - n_0)\,.
\label{E-gen:1}
\eqa
The parameters $c_E$ and $r_s$ depend on the interaction potential,
while $b$ is a universal coefficient.
The chemical potential $\mu$ is determined by the condition
$\partial {\cal F}/\partial v =0$, which is equivalent to
\bqa
{\partial \ \over \partial n_0} \left( {\cal E} - \mu n \right) = 0.
\eqa
The constant $4/(9\pi^2)$ in the coefficient of the $n_0 a^3$
correction in (\ref{E-gen:1}) is part of the universal coefficient. 
The rest of that coefficient has been
absorbed into the nonuniversal coefficient $c_E$.
This represents a specific definition of the three-body contact parameter 
$c$ whose motivation will become evident below in Eq.~(\ref{E-gen:2}).

The low-density expansion for the energy density
can be obtained by using (\ref{n0-gen}) to eliminate $n_0$ 
from (\ref{E-gen:1}) in favor of $n$.
Including all terms through order $(na^3)^{3/2}$,
the expression for the energy density is
\bqa
{{\cal E} \over {\cal E}_{\rm umf}} &=&  
1 + {32 \over 15 \pi} \sqrt {16 \pi n a^3}
+ \left [ c_E + {4 \pi - 3 \sqrt 3 \over 6 \pi} 
	\log (16 \pi n a^3) \right ] (16 \pi n a^3) 
\nonumber
\\
&&  \hspace{1cm}
+ \left( {16 \over \pi} \left[ c_E 
	+ {4 \pi - 3 \sqrt 3 \over 6 \pi} 
		\log (16 \pi n a^3) \right]
	- {16 \over 15 \pi} {r_s \over a} 
	+ b' \right) 
	(16 \pi n a^3)^{3/2} \,,
\label{E-gen:2}
\eqa
where $b'=b -{4\over3\pi}d - {8\over9\pi^3}$.

We have chosen to define the three-body contact parameter $c_E$
so that it includes all the universal terms under the logarithm 
$\log (16 \pi n a^3)$ in the $n a^3$ term in (\ref{E-gen:2}).
This is a perfectly acceptable definition of 
the three-body contact parameter, because
it can be related to any other definition by a well-defined calculation.
For example, dimensional regularization and minimal subtraction
can be used to define a three-body coefficient $c_{MS}(\kappa)$ 
that depends on an arbitrary renormalization scale $\kappa$ \cite{BrN99}.
The two-body and three-body coupling constants $g$ and $g_3(\kappa)$ 
in Ref.~\cite{BrN99} are equal to $A_0$ and $B_0$ up to factors of $\hbar$.
The energy density was calculated to order $n a^3$ 
using dimensional regularization and minimal subtraction in 
Ref.~\cite{BrN99}.   
By equating the coefficient of the $(16 \pi n a^3)$ term
in the expression (\ref{E-gen:2}) with the corresponding coefficient in 
Ref.~\cite{BrN99}, we find
\beq
c_E + {4 \pi - 3 \sqrt 3 \over 6 \pi} 
	\log(16 \pi n a^3)
\;=\; c_{\rm MS}(\kappa) 
+ {4 \pi - 3 \sqrt 3 \over 6 \pi}
	\left[\log {16 \pi n a \over \kappa^2} + 0.80 \right] \,.
\eeq
Thus the parameter $c_E$
defined by the expression (\ref{E-gen:2}) is equal to the running parameter
$c_{\rm MS}(\kappa)$ at a scale of order $1/a$:
\beq
c_E \;=\; c_{\rm MS}(\kappa = 1.5/a) \,.
\eeq

\section{Diffusion Monte Carlo Results}
\label{sec4}

In this section, we present the diffusion Monte Carlo results of Giorgini,
Boronat, and Casulleras \cite{GBC99} for the
energy and the condensate fraction of the homogeneous Bose gas.  
We also compute the 
effective ranges for each of the model potentials they considered.  

\subsection{Model potentials}

In the low-density limit, a homogeneous Bose gas exhibits universal behavior
that is determined completely by the S-wave scattering length $a$.  However, at
any nonzero density, its properties will also depend at some level on aspects
of the interactions between atoms other than the scattering length.  These
nonuniversal effects scale as higher powers of $n$ and become negligible in the
low-density limit.

If Ref. \cite{GBC99}, the authors considered 4 different two-body potentials 
as models for the interactions between the bosons.  These potentials all 
have the same scattering length $a$.  The potentials are
\begin{itemize}

\item the hard-sphere potential (HS):
\bqa
V^{\rm (HS)}(r) & = & \infty, \hspace{1cm}   r < a\,,
\nonumber
\\
      & = & 0,      \hspace{1.2cm} r > a\,.
\eqa

\item two soft-sphere potentials (SS-5 and SS-10):
\bqa
V^{\rm (SS)}(r) & = & V_0, \hspace{1cm}   r < R\,,
\nonumber
\\
      & = & 0,   \hspace{1.2cm} r > R \,.
\eqa
The radii are $R = 5 \, a$  for SS-5 and $R = 10 \, a$ for SS-10.
The heights of the two potentials are 
$V_0 = 0.031543 \, \hbar^2 / (ma^2)$ for SS-5 and 
$V_0 = 0.0034084 \, \hbar^2/(ma^2)$ for SS-10. 

\item a hard-core square-well potential (HCSW):
\bqa
V^{\rm (HCSW)}(r) & = & + \infty, \hspace{1cm}   r < R_c\,,
\nonumber
\\
      & = & - V_0,    \hspace{1cm}   R_c < r < R\,,
\nonumber
\\
      & = & 0,        \hspace{1.6cm} r > R \,.
\eqa
The outer radii of the hard core and the square well are $R_c = a/50$ 
and $R = a/10$, respectively.   The depth of the square well is 
$V_0 = 412.815 \, \hbar^2 / (ma^2)$.  
The potential supports a single two-body bound state with energy 
$- 1.13249 \, \hbar^2 /(ma^2)$. 

\end{itemize}

Nonuniversal effects in the Bose gas depend on aspects of the interactions
between the bosons other than the S-wave scattering length.  One of the most
important is the effective range for S-wave scattering $r_s$,
which is defined by the low-momentum expansion (\ref{ERexp}) of the 
phase shift for atom-atom scattering.
The effective range for the hard-sphere potential is $r_s = 2a/3$. 
For the soft-sphere potentials, the S-wave scattering length is
\beq
a = R  - {\tanh(k_0 R) \over k_0} \,,
\eeq
where $k_0 = \sqrt{m V_0/\hbar^2}$.  The effective range  is
\beq
r_s = R - {R^3 \over 3 a^2} + {1 \over k_0^2 a} \,.
\eeq
Its numerical value is 
$r_s =  -4.96372 \, a$ for SS-5 and $r_s = -29.936 \, a$ for SS-10. 
For the HCSW potential, the S-wave scattering length is
\beq
a = R  - {\tan[k_0 (R- R_c)] \over k_0} \,.
\eeq
The effective range is
\beq
r_s = R - {R^3 \over 3 a^2} + {R_c(R-a)^2 \over a^2} 
	+ {R_c - a \over k_0^2 a^2} \,.
\eeq
Its numerical value is $r_s = 0.113493 \, a$. 

\subsection{Energy and Condensate Fraction}

In Ref. \cite{GBC99}, the authors used the diffusion Monte Carlo 
method to calculate the properties of a homogeneous gas of bosons 
interacting through each of the four model potentials.  
In the  case of the potentials HS, SS-5, and SS-10, the
homogeneous gas is the ground state of the system.  In the case of the
potential HCSW, the homogeneous gas is a metastable state 
in which there are no two-body bound states.

\begin{table}
\begin{tabular}{cccccc}
$na^3$ & HS & SS-5 & SS-10 & HCSW  \\
\tableline
$1 \times 10^{-6}$ & 1.00427(80) & 1.00427(80) & 1.00427(80) & 1.00427(80) \\
$5 \times 10^{-6}$ & 1.00952(16) & & & \\
$8 \times 10^{-6}$ & 1.01262(99) & & 1.01163(99) & \\
$1 \times 10^{-5}$ & 1.01382(80) & 1.01382(80) & 1.01302(80) & 1.0162(16)  \\ 
$1.2 \times 10^{-5}$ & 1.01528(66) & & 1.01395(66) & \\
$5 \times 10^{-5}$ & 1.02957(48) & & & \\
$1 \times 10^{-4}$ & 1.04326(80) & 1.04167(80) & 1.03689(80) & 1.04565(80) \\ 
$5 \times 10^{-4}$ & 1.09499(64) & & & \\
$1 \times 10^{-3}$ & 1.1332(16)  & 1.11011(80) & 1.07907(80) & 1.1380(40)  \\  
$5 \times 10^{-3}$ & 1.29775(95) & & & \\
$1 \times 10^{-2}$ & 1.42921(80) & & & \\
$5 \times 10^{-2}$ & 2.1295(16)  & & & \\
$1 \times 10^{-1}$ & 2.8855(56)  & & & \\
0.166 & 3.9597(96) & & & \\
0.244 & 5.446(33) & & & \\
\end{tabular}
\caption{Energy normalized to the universal mean-field energy
	for the HS, SS-5, SS-10, and HCSW  potentials 
	from the diffusion Monte Carlo calculations of
        Ref. \protect\cite{GBC99}. The statistical errors are given 
        in parentheses. 
}
\label{tab:E}
\end{table}                         

They calculated the energy per particle $E/N$ 
and the condensate fraction $N_0 / N$ as functions of $na^3$, 
with $na^3$ ranging over several orders of
magnitude from $10^{-6}$ to 0.244.  Their results are reproduced
in Tables \ref{tab:E} and \ref{tab:N}.
The condensate fraction was not calculated for the HCSW potential.  
Most of the values in Table \ref{tab:E} were given in a table
in Ref.~\cite{GBC99}. The energy values for $na^3 =8\times 10^{-6}$
and $na^3 =1.2\times 10^{-5}$ and all the condensate values in Table
\ref{tab:N} were provided to us by the authors.
For convenience, we have normalized $E/N$ to the universal mean-field result 
$E_{\rm umf}/N = (2 \pi \hbar^2 /m) a n$.
The errors in parentheses are the statistical errors from the diffusion
Monte Carlo.  The number of particles in the simulation box 
was $N = 500$, which the authors claimed is large enough
that finite-size effects are well below the statistical errors.

\begin{table}
\begin{tabular}{ccccc}
$na^3$ & HS & SS-5 & SS-10 \\
\tableline
$1 \times 10^{-6}$ & 0.998(1) & 0.997(1) & 0.997(2) \\	
$5 \times 10^{-6}$ & 0.997(1) & & \\
$1 \times 10^{-5}$ & 0.996(2) & 0.997(2) & 0.996(1)\\  
$5 \times 10^{-5}$ & 0.991(2) & & \\
$1 \times 10^{-4}$ & 0.987(3) & 0.989(2) & 0.992(1) \\
$5 \times 10^{-4}$ & 0.967(3) & & \\
$1 \times 10^{-3}$ & 0.948(5) & 0.971(3) & 0.986(1)\\ 
$5 \times 10^{-3}$ & 0.876(6) & & \\ 
$1 \times 10^{-2}$ & 0.803(3) & & \\ 
$5 \times 10^{-2}$ & 0.501(5) & & \\ 
$1 \times 10^{-1}$ & 0.277(3) & & \\ 
0.166              & 0.109(4) & & \\ 
0.244              & 0.031(4) & & \\
\end{tabular}
\caption{Condensate fractions for the HS, SS-5, and SS-10 potentials
	from the diffusion Monte Carlo calculations of
	Ref. \protect\cite{GBC99}. The statistical errors are given in 
	parentheses. 
}
\label{tab:N}
\end{table}                                  

\section{Analysis}
\label{sec5}

In this section, we obtain crude determinations of the three-body contact 
parameter $c_E$ defined by (\ref{E-gen:2})
for each of the interaction potentials 
by analyzing the diffusion Monte Carlo data for the energy.
We also use the diffusion Monte Carlo data for the HS potential
to obtain a crude determination of the universal coefficient $d'$
in the low-density expansion (\ref{n0-gen}) for the condensate fraction.

\subsection{Three-body contact parameters}

In Section \ref{sec4}, we have calculated the effective range
$r_s^{\rm (pot)}$ for each of the 
four model potentials used in Ref.~\cite{GBC99}.
As a consequence, the only unknowns in the expression (\ref{E-gen:2}) 
for the energy density
are the three-body coefficient $c_E^{\rm (pot)}$,
which depends on the interaction potential,
and the universal coefficient $b'$. At small densities, the contribution
of $b'$ is suppressed by a power of $\sqrt{na^3}$ 
compared to the contribution of $c_E^{\rm (pot)}$.
We define a function $C^{\rm (pot)}(na^3 )$ that in the limit $na^3 \to 0$
approaches $c_E^{\rm (pot)}$ via
\bqa
\label{C-pot}
C^{\rm (pot)}(na^3 )&=&
{1 \over 16 \pi n a^3}
\left[ {{\cal E}^{\rm (pot)} \over {\cal E}_{\rm umf}}- 
1 - {32 \over 15 \pi} \sqrt {16 \pi n a^3}
	- {4 \pi - 3 \sqrt 3 \over 6 \pi} 
	(16 \pi n a^3)\log (16 \pi n a^3)  \right]
\,.
\eqa
The approach to the limiting value $c^{\rm (pot)}$ 
as $na^3 \to 0$ is given by
\bqa
C^{\rm (pot)} &\longrightarrow&
c_E^{\rm (pot)} 
+ \left( {16 \over \pi} \left[   c_E^{\rm (pot)} 
	+ {4 \pi - 3 \sqrt 3 \over 6 \pi} 
		\log (16 \pi n a^3) \right] 
	-{16 \over 15 \pi} {r_s^{\rm (pot)} \over a} 
	+ b' \right) 
	(16 \pi n a^3)^{1/2}\,.
\label{C-pot:2}
\eqa
For sufficiently small values of $na^3$,
the approach of the function $C^{\rm (pot)}$ to the constant
value $c_E^{\rm (pot)}$ should be dominated by the logarithmic 
term in (\ref{C-pot:2}).

\begin{figure}[ht]
\bigskip
\epsfxsize=12.cm
\centerline{\epsffile{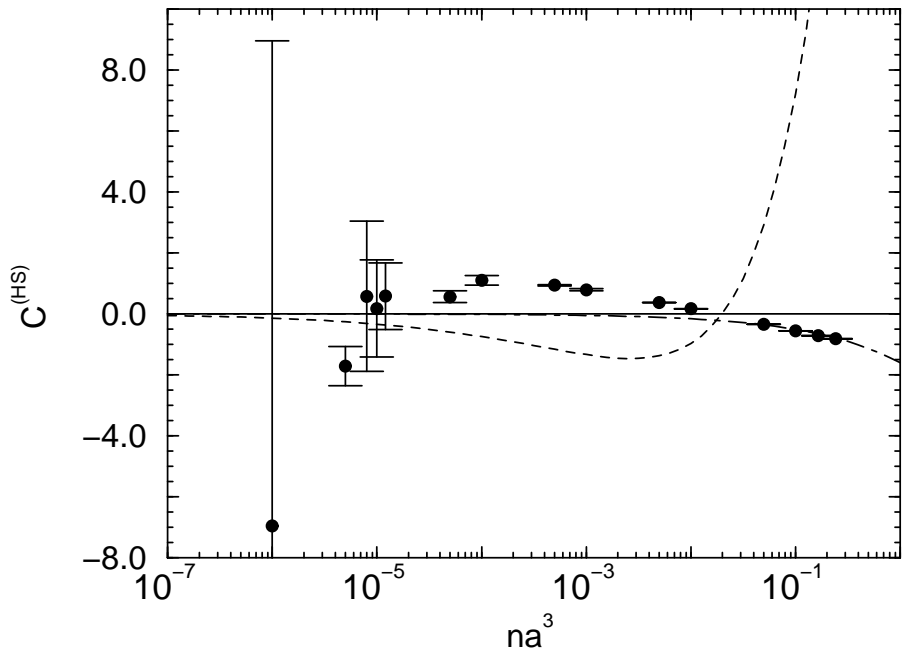}}
\bigskip
\caption{The value of $C^{\rm (HS)}$ for the data points
from Table \ref{tab:E}.
The dot-dashed curve shows the corrections from the effective range
term. The dashed curve in shows the corrections from the
logarithmic term at the next order.}
\label{fig1}
\end{figure}

We first consider the HS potential.
In Fig.~\ref{fig1}, we show the values of $C^{\rm (HS)}$ calculated from
the diffusion Monte Carlo data in Table~\ref{tab:E}.
The curves give some idea of the expected size of the  
corrections from next order in $\sqrt{na^3}$.  
The dashed and dot-dashed curves show the 
corrections from the logarithmic term and the effective range term,
respectively. 
The correction from the $b'$ term in (\ref{C-pot:2}) has the same 
density dependence as the effective range term, 
but its normalization in unknown. 
The logarithmic correction will be the largest 
in the region $na^3 \le 1.2 \times 10^{-5}$ unless 
$|b'| > 14.8$ or $|c_E^{\rm (HS)}|>2.9$.
For $na^3 \le 1.2 \cdot 10^{-5}$,
the statistical errors in the data are larger than the expected size
of the corrections from higher orders.
We can therefore obtain a rough determination of $c_E^{\rm (HS)}$ by
averaging $C^{\rm (HS)}$ at the 5 density values
$na^3 \le 1.2\times 10^{-5}$ weighted by the inverse-squares 
of the statistical errors.
The lowest data point could equally well be omitted 
because of its large error bar. 
The resulting value is
\beq
\label{c-HS}
c_E^{\rm (HS)} \;=\; -0.9 \pm 0.5\,.
\eeq
The error was estimated by adding the inverses of the statistical 
errors in quadrature and then taking the reciprocal. 
The value (\ref{c-HS}) is dominated by the data point at 
$na^3 = 5 \times 10^{-6}$, because it has such a small statistical error.
The systematic error from the $\sqrt{na^3}$ corrections in (\ref{C-pot:2})
is expected to be smaller than the magnitude of the logarithmic 
correction in (\ref{C-pot:2}), which is 0.26 at $na^3 = 5 \times 10^{-6}$. 

We also performed a global $\chi^2$-fit of the data using
(\ref{E-gen:2}) through order $na^3$. We assigned a 
theoretical error $\pm (16\pi na^3)^{3/2}$ to quantify the error
from leaving out the unknown higher order terms in the expansion. 
We computed the $\chi^2$ for the difference between the data 
in Table~\ref{tab:E} and the theoretical expression (\ref{E-gen:2})
through order  $na^3$ using the error obtained by adding the theoretical 
error and the statistical error from Table~\ref{tab:E} in quadrature.
The preferred value is $c_E^{\rm (HS)} \;=\; 0.8$.
In this case, the data points at higher density favor a positive
value for $c_E^{\rm (HS)}$ in spite of their larger theoretical error.

\begin{figure}[ht]
\bigskip
\epsfxsize=12.cm
\centerline{\epsffile{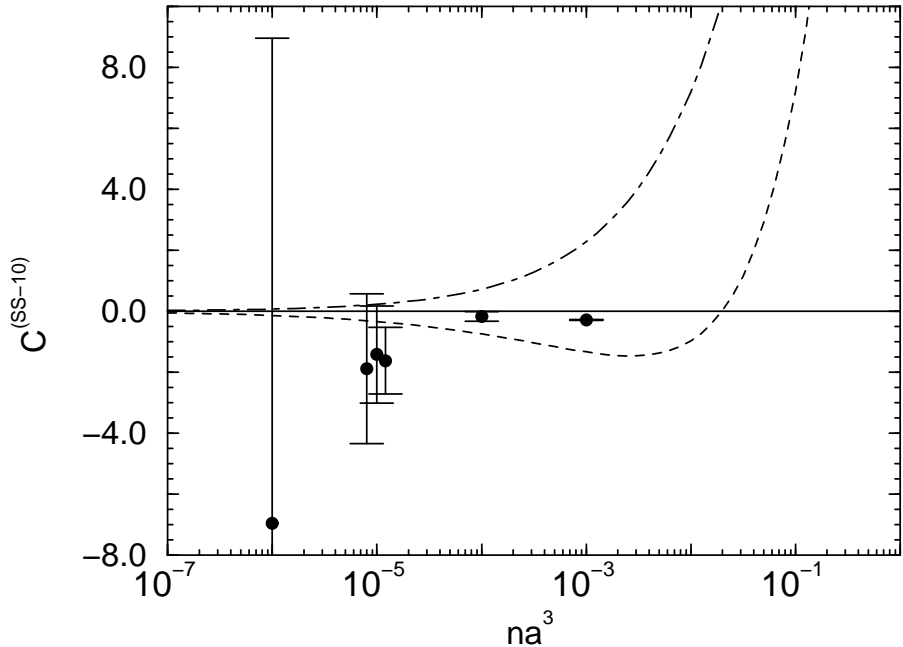}}
\bigskip
\caption{The values of $C^{\rm (SS-10)}$ for the data points from
Table \ref{tab:E}. The curves are as in Fig.~\ref{fig1}.}
\label{fig2}
\end{figure}

We next consider the SS-10 potential, 
for which there are six data points in Table~\ref{tab:E}.
In Fig.~\ref{fig2}, we show the values of $C^{\rm (SS-10)}$ calculated from
the diffusion Monte Carlo data for the SS-10 potential
in Table~\ref{tab:E}.
The curves are the same as in Fig.~\ref{fig1}, except that
the effective range correction is much larger,
because $r_s^{\rm (SS-10)}$ is larger than $r_s^{\rm (HS)}$ 
by about a factor of 45. 
For $na^3 \le 1.2 \cdot 10^{-5}$,
the statistical errors in the data are larger than the expected size
of the corrections from higher orders.
We can therefore obtain a rough determination of $c^{\rm (SS-10)}$ by
averaging $C^{\rm (SS-10)}$ at the 4 density values
$na^3 \le 1.2\times 10^{-5}$ weighted by the inverse-squares 
of the statistical errors.
The resulting value is
\beq
\label{c-SS10}
c_E^{\rm (SS-10)} \;=\; -1.6 \pm 0.8\,.
\eeq

\begin{figure}[ht]
\bigskip
\epsfxsize=14.cm
\centerline{\epsffile{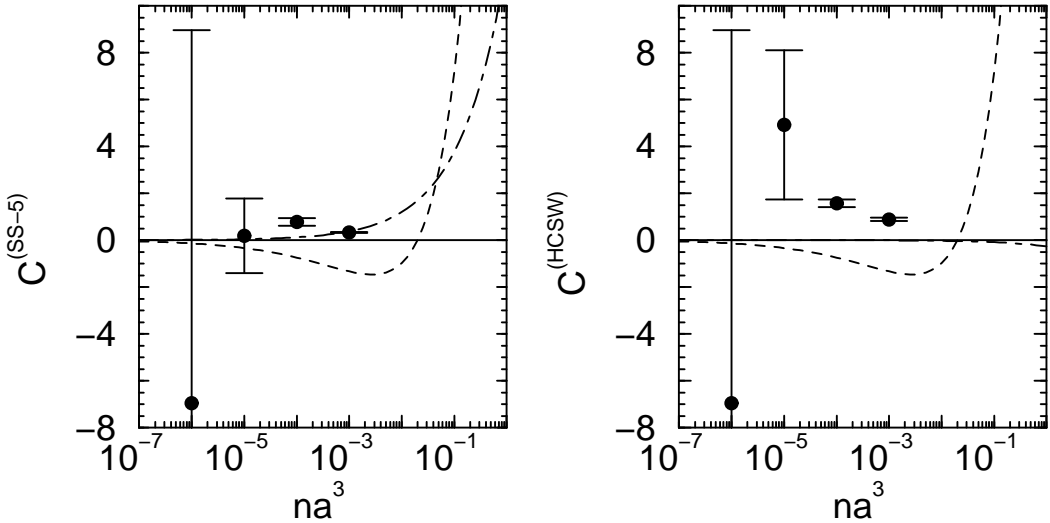}}
\bigskip
\caption{The values of $C^{\rm (SS-5)}$ (left plot) and 
$C^{\rm (HCSW)}$ (right plot) for the data points from
Table \ref{tab:E}. The curves are as in Fig.~\ref{fig1}.}
\label{fig3}
\end{figure}

For both the SS-5 and HCSW potential,
there are only four data points in Table~\ref{tab:E}.
In Fig.~\ref{fig3}, we show the corresponding values of 
$C^{(\rm SS-5)}$ and $C^{(\rm HCSW)}$.
The expected error from higher orders is smaller than the statistical 
error for the 2 data points with $na^3 \le 10^{-5}$.
If we use the value of $C^{\rm (pot)}$ at $na^3 = 10^{-5}$ 
as an estimate for $c_E^{\rm (pot)}$, we obtain
$c_E^{\rm (SS-5)}= 0.2 \pm 1.6$ and $c_E^{\rm(HCSW)}= 4.9 \pm 3.2$.

We have given a rough determinations of the three-body contact parameter
$c_E^{\rm (pot)}$ for each of the potentials HS, SS-10, SS-5, and HCSW.
The error bars are large enough that we cannot identify a deviation 
from universality in the three-body coefficients. 
More accurate determinations would require higher statistics
in the diffusion Monte Carlo calculations.
It might also require increasing the number of particles 
in the simulation box in order to keep finite-size effects 
smaller than the statistical errors.
Ideally, one would like to have high enough statistics at several 
values of $na^3$ between $10^{-4}$ and $10^{-6}$ so that the rate of 
approach of $C^{\rm (pot)}(na^3)$ to the constant value $c_E^{\rm (pot)}$
could also be used to determine the universal coefficient $b'$ 
in (\ref{C-pot:2}).

\subsection{Universal 2nd-order correction to the condensate fraction}

In the expression (\ref{n0-gen}) for the condensate fraction,
the only unknowns are the three-body coefficient $c_E^{\rm (pot)}$
and the universal coefficients $d'$ and $e'$. 
We define a function $D^{\rm (pot)}(na^3 )$ that approaches the universal
coefficient $d'$ as the density approaches zero by
\bqa
D^{\rm (pot)}(na^3) & =& 
- {1 \over 16 \pi n a^3} 
\left[ \left({n_0 \over n}\right)^{\rm (pot)} - 1
	+ {2 \over 3 \pi} \sqrt {16 \pi n a^3} 
	\right] \,.
\label{D-pot}
\eqa
The approach to the limiting value $d'$ as $na^3 \to 0$ is given by
\bqa
D^{\rm (pot)} &\longrightarrow& d' 
+ \left( {3 \over \pi} \left[  c_E^{\rm (pot)} 
	+ {4 \pi - 3 \sqrt 3 \over 6 \pi} 
		\log (16 \pi n a^3) \right]
	- {1 \over 2 \pi} {r_s^{\rm (pot)} \over a} 
	+ e' \right)
	(16 \pi n a^3)^{1/2} \,.
\label{D-pot:2}
\eqa

\begin{figure}[htb]
\bigskip
\epsfxsize=12.cm
\centerline{\epsffile{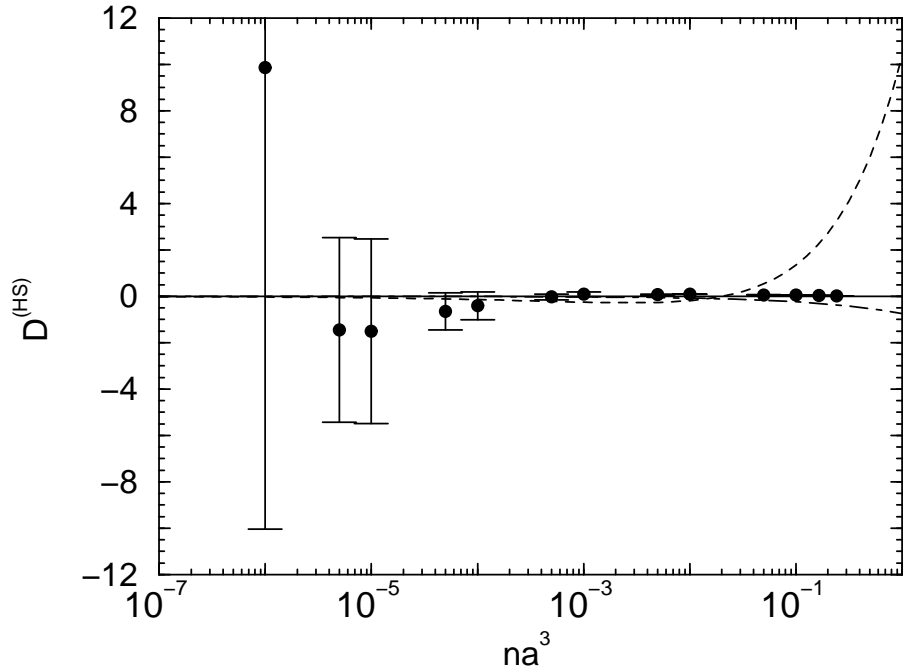}}
\bigskip
\caption{The values of $D^{\rm(HS)}$ calculated from the diffusion Monte
         Carlo data in  Table~\ref{tab:N}. The dashed line shows the 
         corrections from the logarithmic term while the dash-dotted
         line shows the corrections from the effective range.}
\label{fig4}
\end{figure}

In Fig. \ref{fig4}, we show the values of $D^{\rm(HS)}$ calculated from the 
diffusion Monte Carlo data for the HS potential in  Table~\ref{tab:N}.
The curves give some idea of the expected size of the  
corrections from next order in $\sqrt{na^3}$.  
The dashed and dot-dashed curves show the 
corrections from the logarithmic term and the effective range term,
respectively.
The correction from the $e'$ term in (\ref{D-pot:2}) has the same 
density dependence as the effective range term,
but its normalization in unknown. 
 The logarithmic correction should be the largest in the region 
$na^3 \le 10^{-4}$, unless $|e'| > 2.0$ or $|c_E^{\rm(HS)}|>2.1$.
For $na^3 \le 10^{-4}$,
the statistical errors in the data are larger than the expected size
of the corrections from higher orders.
We can therefore obtain a rough determination of $d'$ by
averaging $D^{\rm (HS)}$ at the 5 density values
$na^3 \le 10^{-4}$ weighted by the inverse-squares 
of the statistical errors.  The resulting value is
\beq
d' =-0.5\pm 0.5\,.
\label{d-HS}
\eeq
The systematic error from the $\sqrt{na^3}$  corrections in (\ref{D-pot:2})
is expected to be smaller than the size of the logarithmic correction
in (\ref{D-pot:2}), which is 0.14. 
We have also determined $d'$ by using a global $\chi^2$-fit 
to the data of Eq. (\ref{n-gen}) through order $na^3$, using
a theoretical error of $\pm(16 \pi n a^3)^{3/2}$
that is added  in quadrature to the statistical error.
The preferred value is $d'=0$, which agrees with Eq. (\ref{d-HS}) 
to within the error bars.

\begin{figure}[htb]
\bigskip
\epsfxsize=14.cm
\centerline{\epsffile{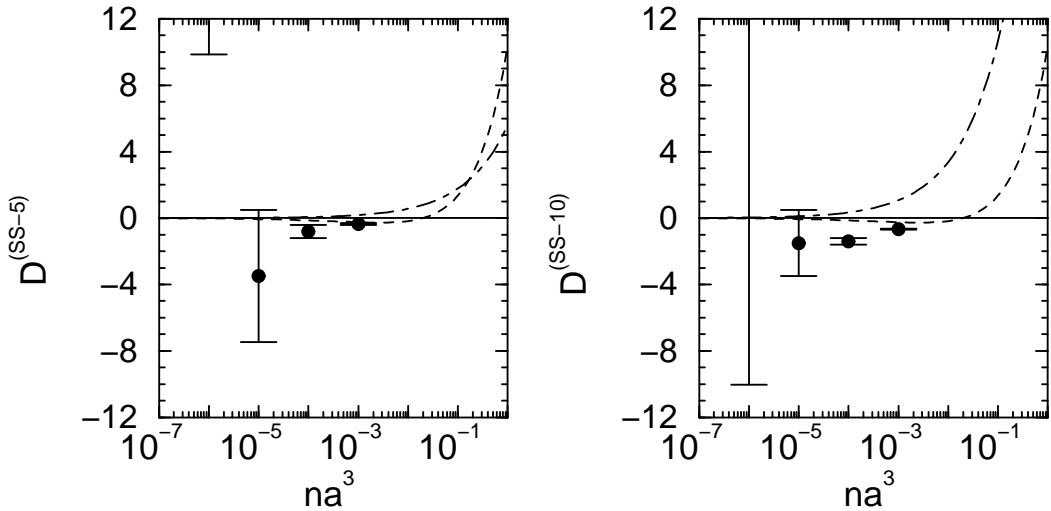}}
\bigskip
\caption{The values of $D^{\rm (SS-5)}$ (left plot) and $D^{\rm (SS-10)}$
	(right plot)  calculated from the diffusion Monte
         Carlo data in  Table~\ref{tab:N}. 
	 The curves are as in Fig.~\ref{fig4}.
         The first data point at $na^3=10^{-6}$ 
         is $29.8 \pm 19.9$ for the SS-5 potential
         and $29.8 \pm 39.8$ for the SS-10 potential.}
\label{fig5}
\end{figure}

For both the SS-5 and SS-10 potentials, there are only 4 data points for the
condensate fraction in Table~\ref{tab:N}.
The condensate fraction was not calculated 
in Ref.~\cite{GBC99} for the HCSW potential.
In Fig. \ref{fig5}, we show the values of $D^{\rm (SS-5)}$ and
$D^{\rm (SS-10)}$ calculated from the 
diffusion Monte Carlo data in Table~\ref{tab:N}.
The curves are the same as in Fig.~\ref{fig2}, 
except for the different value of the effective range.
The expected error from higher orders is smaller than the statistical 
error for the 2 data points with $na^3 \le 10^{-5}$.
If we use the value of $D^{\rm (pot)}$ at $na^3 = 10^{-5}$ 
as an estimate for $d'$, we obtain
$d' = -3.5 \pm 4.0$ from the SS-5 potential 
and $d' =-1.5 \pm 2.0$ from the SS-10 potential. 
These values are consistent within errors with the value
(\ref{d-HS}) obtained from the condensate fraction 
for the hard-shere potential.

\section{Summary}
\label{sec6}

Effective field theory is an ideal tool for the theoretical study 
of nonuniversal effects in the homogeneous Bose gas.
The central idea of effective field theory
is to exploit the  separation of scales in a physical system. 
In a low-density Bose gas, 
the important length scales include the interparticle spacing,
which is proportional to $n^{-1/3}$, and the 
coherence length, which is proportional to $(n a)^{-1/2}$.
These are both much larger than the physical size of the atoms,
and in typical experiments they are also much larger than the scattering
length $a$.
The number of atoms per cubic scattering length $n a^3$ 
is therefore a small parameter, and it allows for a controlled, 
perturbative expansion of low-density observables
in $\sqrt{n a^3}$. This systematic expansion is conveniently 
implemented using effective field theory. 
In low-energy observables, all the sensitivity to
the details of the interactions between atoms
at short distances is captured in low-energy constants, such as
the scattering length, the effective range,
and the effective three-body contact interaction.
Universal results that depend only on the scattering length
can also be obtained using more traditional many-body methods,
but the calculations can be simplified using effective field theory. 
However, the real power of effective field theory becomes apparent
in the systematic calculation of nonuniversal corrections. 

Effective field theory can be used to easily determine at what order
in the low-density expansion various nonuniversal effects will enter.
The scattering length is associated with the $(\psi^*\psi)^2$
term in the local effective Lagrangian (\ref{Lloc}).
The low-energy parameters associated with nonuniversal effects 
can be identified with terms in the effective Lagrangian
that are higher order in $\psi^*\psi$ or in $\nabla$.
In general, there are infinitely many such terms, but 
only a finite number contribute to any observable 
at a given order in the low-density expansion.
The order in $\sqrt{n}$ at which a given term contributes 
is determined by the powers of $\psi^*\psi$ 
and $\nabla$ and by the order in the loop expansion
at which it first enters. 
For example, if we consider the energy density,
every power of $\psi^*\psi$ contributes a factor of $n$. 
Every power of $\nabla$ contributes a factor of $\sqrt{na}$.
Finally there is a quantum suppression factor of $\sqrt{na^3}$
associated with each order in the loop expansion.
For example, the three-body contact term $(\psi^*\psi)^3$
gives a mean-field contribution proportional to $n^3$.
The effective range term $[\nabla (\psi^*\psi)]^2$
does not contribute in the mean-field approximation, 
because the gradients vanish for a homogeneous gas. 
Its first contribution is at the semiclassical (one-loop) order and 
is proportional to 
$\sqrt{na^3} (\sqrt{na})^2 n^2 \sim n^{7/2}$.
Thus the leading nonuniversal contribution comes from the three-body contact
interaction.  The effective range enters at one order higher in
$\sqrt{n}$. At order $n^4$, there is also a nonuniversal contribution 
to the energy density associated with the $(\psi^*\psi)^4$ term 
in the effective Lagrangian.

We have calculated the leading 
and next-to-leading nonuniversal corrections
to the energy density and the leading nonuniversal corrections
to the condensate fraction for a homogeneous Bose gas.
These nonuniversal
corrections are determined by the effective range $r_s$ and 
the three-body contact parameter $c$. 
These parameters are in general different for different potentials, 
even if they have the same scattering length $a$.
For a given potential, the effective range can be extracted 
from the amplitude for $2\to 2$ scattering in the standard
way. The parameter $c$ could in principle be determined from a three-body 
observable, such as the amplitude for $3\to 3$ scattering in the vacuum.
However, it can equally well be obtained from a low-density observable.
In fact, we chose to define the  
parameter $c_E$ using the form of the low-density expansion 
for the energy density in (\ref{E-gen:2}).

We analyzed the nonuniversal effects in the diffusion Monte Carlo
results of Giorgini, Boronat, and Casulleras \cite{GBC99}
for a homogeneous Bose gas with four different interaction potentials.
We attempted to extract the unknown three-body coefficient 
$c_E$ from the diffusion Monte Carlo data for the
energy density. 
We found that the statistical errors at small values of $na^3$
allowed only a crude determination of this parameter.
Our values for the hard-sphere and SS-10 potentials are given in
(\ref{c-HS}) and (\ref{c-SS10}), respectively.  
Within the error bars, we were unable to see a deviation from universality
in the three-body coefficients. To determine $c_E$ precisely enough 
to see the nonuniversality, one would need higher statistics 
at several densities in the range $10^{-4}< na^3<10^{-6}$. 

We also attempted to extract from the diffusion Monte Carlo data
the universal coefficient $d'$
in the low-density expansion (\ref{n0-gen}) for the condensate fraction.
The value obtained from the data for the hard-shere potential is given in
(\ref{d-HS}).  Again we found that the statistical errors 
at small values of $na^3$ allowed only
a crude determination of this coefficient.
In order to get a useful determination of $d'$, 
one would need higher statistics 
at several densities in the range $10^{-3} < na^3 < 10^{-5}$.
The coefficient $d'$ can also be calculated from Feynman diagrams 
using the methods used in Ref.~\cite{BrN99} to compute the 
$na^3$ corrections to the energy density.  
Such a calculation would provide a
useful check on the accuracy of diffusion Monte Carlo results.

The achievement of Bose-Einstein condensates in atomic vapors 
has opened up the possibility for experimental study
of Bose gases.  As the experiments become more and more precise, 
it will be important to understand nonuniversal effects.
Effective field theory provides a perturbative framework 
within which nonuniversal effects can be systematically calculated
in the low density limit in terms of a few low-energy parameters.
These effects can also be calculated nonperturbatively using
diffusion Monte Carlo simulations.  More extensive 
diffusion Monte Carlo calculations would be extremely valuable 
for testing the understanding of nonuniversal effects provided by 
effective field theory.

\acknowledgements
We thank S.~Giorgini, J.~Boronat, and J.~Casulleras for providing
us with the results of their diffusion Monte Carlo simulations.
We also thank R.J.~Furnstahl for useful discussions.
HWH thanks the Institute for Nuclear Theory at the University of Washington
for its hospitality and partial support during completion of this work.
This research was supported in part by DOE grant DE-FG02-91-ER4069 
and by NSF grant PHY-9800964.

\end{document}